\documentclass{ws-procs9x6}
\usepackage{graphicx}
\usepackage{amsmath}
\usepackage{amsfonts}
\usepackage{amssymb}
\begin{document}

\title{Using Reactors to Measure $\theta_{13}$}
\author{M. H. Shaevitz and J. M. Link\\Columbia University,\\Dept. of Physics \\New York, NY 10027, USA}
\maketitle

\abstracts{ A next-generation neutrino oscillation experiment
using reactor neutrinos could give important information on the
size of mixing angle $\theta_{13}$. The motivation and goals for a
new reactor measurement are discussed in the context of other
measurements using off-axis accelerator neutrino beams. The reactor
measurements give a clean measure of the mixing angle without
ambiguities associated with the size of the other mixing angles,
matter effects, and effects due to CP violation. The key question
is whether a next-generation experiment can reach the needed
sensitivity goals to make a measurement for $\sin^{2}2\theta_{13}$
at the 0.01 level. The limiting factors associated with a reactor
disappearance measurement are described with some ideas of how
sensitivities can be improved. Examples of possible experimental
setups are presented and compared with respect to cost and
sensitivity. }

\section{Motivation and Goals of a Next-Generation Reactor Oscillation Experiment}

Information on the masses and mixing angles in the neutrino sector is growing
rapidly and the current program of experiments will map out the parameters
associated with the solar, atmospheric, and LSND signal. With the recent
confirmation by KamLAND and isolation of the $\Delta m_{solar}^{2}$ in the LMA
region, the emphasis of many future neutrino oscillation experiments is
turning to measuring the last mixing angle, $\theta_{13}$ and obtaining better
precision on $\Delta m_{solar}^{2}$ and $\Delta m_{atm}^{2}$ (along with
checking LSND).

A road map for future, worldwide neutrino oscillation measurements can be
considered as connected stages. Stage 0 includes the current program with K2K,
CNGS, and NuMI/Minos probing the $\Delta m_{atm}^{2}$ region with the goal of
measuring $\Delta m_{23}^{2}$ to about 10\%. MiniBooNE over this time period
will make a definitive check of the LSND anomaly and measure the associated
$\Delta m^{2}$ and mixing if a signal is observed.

A next step, Stage 1, would have the goal of measuring or limiting the value
of $\theta_{13}$. At this stage, experiments could possibly see the first
indications of CP violation and matter effects if $\theta_{13}$ is large
enough. For $\theta_{13}$, the NuMI/Minos on-axis experiment has sensitivity
for $\sin^{2}2\theta_{13}>0.06$ at 90\% CL. Better sensitivity experiments are
being proposed for this stage including the NuMI and JHF off-axis experiments
along with two detector reactor experiments. The combination of off-axis and
reactor measurements is a powerful tool for isolating the physics. In the end,
these experiments need to provide information on $\sin^{2}2\theta_{13}>0.01$
at the 3$\sigma$ measurement level as a prerequisite for building the
expensive Stage 2 experiments.

The goal of Stage 2 would be to observe CP violation and matter effects. One
component of this stage will be high intensity neutrino sources combined with
large detectors ($>500$ ktons) at long baselines. Due to ambiguities in how
the various physics processes manifest themselves, the program is best
accomplished using a combination of high statistics neutrino and antineutrino
measurements at various baselines combined with high statistics reactor
measurements. If Stage 2 is successful, a Stage 3 would use a muon storage
ring, neutrino factory to map out CP violation in the neutrino sector and make
measurements with a precision one to two orders of magnitude better than Stage 2.

For measuring $\theta_{13}$, reactor measurements are an important ingredient
if the required sensitivity can be reached. Reactors are a very high flux
source of antineutrinos and have been used in the past for several neutrino
oscillation searches and measurements (Bugey, CHOOZ, Palo Verde, and KamLAND).
Currently, several groups are considering new reactor oscillation experiments
with the primary goal of improved sensitivity to the MNS mixing angle,
$\theta_{13}$. To improve sensitivity, the new experiments will use a
comparison of detectors at various distances from the reactor thus minimizing
the uncertainties due to the reactor neutrino flux.

\section{Appearance versus Disappearance Measurements}

An appearance measurements of $\theta_{13}$ can be accomplished by observing
an excess of $\nu_{e}$ events in fairly pure $\nu_{\mu}$ beam. The measurement
is difficult since the signal is a small number of $\nu_{e}$ events over a
comparable background. The proposed new JHF-SuperK\cite{JHF-SK} and NuMI
off-axis\cite{NuMI-off-axis} experiments are to use far detectors placed
off-axis with respect to the neutrino beam direction. Due to the kinematics of
pion decay, the off-axis setup gives a beam with a sharp energy spectrum which
minimizes neutral current $\pi^{0}$ backgrounds and allows the energy to be
tuned to the first oscillation maximum.

\begin{figure}
[ptb]
\begin{center}
\includegraphics[
height=2.3168in,
width=4.5887in
]%
{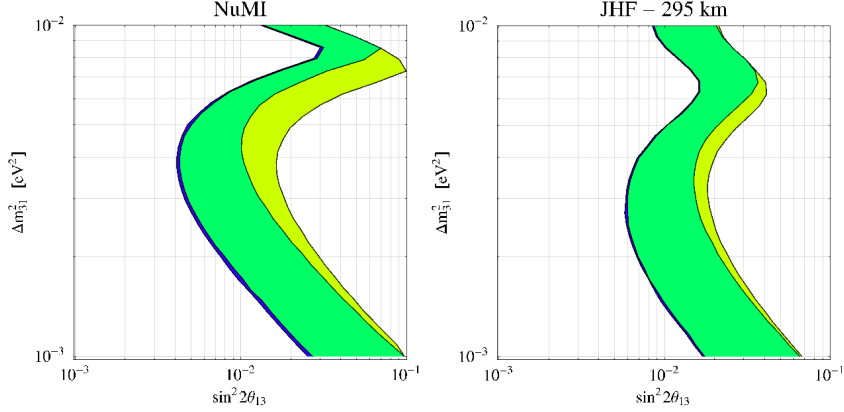}%
\caption{Ambiguity bands for interpreting the $\nu_{\mu}\rightarrow\nu_{e}$
transition probability in terms of $\sin^{2}\theta_{13}$ for the NuMI (712
km), left, and the JHF (295 km), right. The main part of the bands on the left
are due to the value of $\delta$ and that on right from whether $\theta
_{23}<\pi/4$ or $>\pi/4$. (From Ref. 3)}%
\label{appear_sense}%
\end{center}
\end{figure}

The off-axis experiments measure the $\nu_{\mu}\rightarrow\nu_{e}$ transition
probability as given in Eq. \ref{numu-nue} (where $\sin\theta_{23}=\frac
{1\pm\sqrt{1-\sin^{2}2\theta_{23}}}{2}$ and $\Delta_{ij}=\Delta m_{ij}%
^{2}L/(4E)=(m_{i}^{2}-m_{j}^{2})L/(4E)$ ). This transition probability is
mainly proportional to $\sin^{2}2\theta_{13}$ but has ambiguities from the
knowledge of $\sin^{2}\theta_{23}$ as well as matter and CP violation effects.
The ambiguities can enhance or reduce the oscillation probability as shown in
Fig. \ref{appear_sense}\cite{Huber1} where the bands reflect the uncertainties
in $\delta$ and $\theta_{23}$.%

\begin{align}
P\left(  \overset{(-)}{\nu}_{\mu}\rightarrow\overset{(-)}{\nu}_{e}\right)   &
=\sin^{2}\theta_{23}\sin^{2}2\theta_{13}\sin^{2}\Delta_{31}\label{numu-nue}\\
&  \pm\frac{\Delta m_{21}^{2}}{\Delta m_{31}^{2}}\sin2\theta_{13}\sin
\delta_{CP}\cos\theta_{13}\sin2\theta_{12}\sin2\theta_{23}\sin^{3}\Delta
_{31}+...\nonumber
\end{align}%

%TCIMACRO{\FRAME{ftbpFU}{4.5887in}{2.3168in}{0pt}{\Qcb{Ambiguity bands for
%interpreting the $\nu_{\mu}\rightarrow\nu_{e}$ transition probability in terms
%of $\sin^{2}\theta_{13}$ for the NuMI (712 km), left, and the JHF (295 km),
%right. The main part of the bands on the left are due to the value of $\delta$
%and that on right from whether $\theta_{23}<\pi/4$ or $>\pi/4$. (From Ref.
%3)}}{\Qlb{appear_sense}}{offsen.eps}{\special{ language "Scientific Word";
%type "GRAPHIC";  maintain-aspect-ratio TRUE;  display "USEDEF";
%valid_file "F";  width 4.5887in;  height 2.3168in;  depth 0pt;
%original-width 13.0474in;  original-height 6.5561in;  cropleft "0";
%filename 'offsen.eps';file-properties "XNPEU";}}}%
%BeginExpansion
%EndExpansion

On the other hand, a reactor disappearance measurement looks for indications
of a reduced rate of $\overline{\nu}_{e}$ events in a detector at some
distance from the source. The disappearance measurement directly measures
$\sin^{2}2\theta_{13}$ without ambiguities from CP violation and matter
effects.%
\begin{equation}
P(\overline{\nu}_{e}\rightarrow\overline{\nu}_{e})=1-\sin^{2}2\theta_{13}%
\sin^{2}\Delta_{31}-...
\end{equation}
Thus, an unambiguous measurement of $\theta_{13}$ using reactors can be a
powerful tool when combined with off-axis measurements to probe for CP
violation and the neutrino mass hierarchy.\cite{Minakata}$^{,}$\cite{Huber2}
The question is whether a next generation reactor experiment can reach the
required sensitivity. In a disappearance measurement, one needs to be able to
isolate a small change in the overall rate which can be difficult due to
uncertainties in reactor flux, cross sections, and detector efficiencies. As
stated above, sensitivities in the range of $\sin^{2}2\theta_{13}\approx0.01$
should be the goal.

\section{Limiting factors in a reactor disappearance measurement}

Previous reactor disappearance experiments used a single detector at a
distance of about 1 km from the reactor complex. Antineutrinos from the
reactor were detected using the inverse $\beta$-decay reaction followed by
neutron capture on hydrogen or gadolinium.
\begin{align}
\overline{\nu}_{e} + p  &  \rightarrow e^{+} + n\\
&  \hspace{0.9cm}\raisebox{0.5em}{$\mid$}\!\negthickspace\rightarrow\ n +
p(Gd) \rightarrow2.2(8)\text{ MeV}\nonumber
\end{align}
%\begin{align}
%\overline{\nu}_{e}+p  &  \rightarrow e^{+}+n\\
%&  \hookrightarrow n+p(Gd)\rightarrow2.2(8)\text{ MeV}\nonumber
%\end{align}
%
%
%
%
%
The two component coincidence signal of an outgoing positron plus gamma-rays
from the neutron capture is powerful tool to reduce backgrounds and isolate
reactor antineutrino events. The major systematic uncertainty was the 2.8\%
uncertainty associated with the reactor flux.

The CHOOZ experiment\cite{CHOOZ} used a five ton fiducial volume detector
under 300 mwe of shielding at 1 km from two 4.25 GW reactors. The event rate
was $\sim$2.2 events/day/ton with 0.2 to 0.4 background events/day/ton. The
other recent experiment to probe this region was the Palo Verde
experiment\cite{PaloVerde} which used a 12 ton detector under on 32 mwe of
shielding at an average baseline of 850 m from three 3.88 GW reactor. For Palo
Verde the event rate was $\sim$7 events/day/ton over a large background rate
of 2.0 events/day/ton.

Improvements to these previous experiments can be accomplished in several
areas. Higher statistics are needed which demands larger detectors in the 50
ton range and/or larger power reactors. To reduce the dominant reactor flux
spectrum and rate uncertainty, a next generation experiment would need two
detectors at near and far locations. The observed rate in the near detector
can then be used to predict that in the far detector where oscillation effects
are to be probed. Making the near and far detectors identical will reduce
relative efficiency uncertainties. In addition, providing the capability to
move the far detector to the near site allows a direct cross calibration of
the two detectors using the high rates available at the closer distance.
Accurate knowledge of background rates especially in the far site are
necessary. To accomplish the needed uncertainty level demands a combination of
shielding, background measurements, and an excellent veto system.

The spectrum of reactor antineutrinos has a broad distribution peaking near an
energy of 3.5 MeV. For a $\Delta m^{2}=2.5\times10^{-3}$ eV$^{2}$, there is a
broad optimum for the position of the far detector between about 1 and 2 km.
For smaller $\Delta m^{2}\approx1\times10^{-3}$ eV$^{2}$, the sensitivity
degrades by about a factor of two as the optimum position is pushed out toward
3 km.

As an example experiment, we consider two 50 ton detectors located for three
years near a 3 GW reactor with the near detector at 150 m. The statistical
sample in a far detector at 1 to 2 km would range from 23,000 to 92,000 events
leading to a statistical error, $\delta\sin^{2}2\theta_{13}\approx0.003-0.007$
$@$ 90\% CL. Assuming an overburden, of 300 mwe and 0.2 background
events/kton/day gives 9000 background events in the far detector. The
background rate can be measured to 3\% during reactor off periods at a single
reactor site leading to measurement uncertainty of $\delta\sin^{2}2\theta
_{13}\approx0.004$. At multiple reactor sites, there typically is no time when
all reactors are off. Extrapolations using partial shutdowns lead to large
background uncertainties corresponding to $\delta\sin^{2}2\theta_{13}$ in the
$0.01-0.02$ range. As discussed below, the effective background rate can be
substantially ($\times10$) reduced by using an extensive veto system combined
with passive shielding. With such a system, the measurement uncertainty can be
reduced to the level of a single reactor site. The final major uncertainty is
associated with the near to far comparison. With identical detectors, relative
efficiency errors of 1 to 2\% should be obtainable leading to measurement
uncertainties for $\sin^{2}2\theta_{13}$ in the $0.02$ range. If the far
detector can move to the near detector site, a cross calibration could reduce
this uncertainty by a factor 2 to 4 depending on the detailed scenario.

Extrapolating from the previous CHOOZ and KamLAND\cite{KamLAND} detectors, a
next generation detector could be improved in several ways. These detectors
used liquid scintillator with buffer regions to cut down backgrounds from
radioactive decays and cosmic ray muons. Possible improvements include low
activity photomultipliers, an improved veto and shielding system, and
capability to move detectors for cross calibration. Adding gadolinium to
enhance the neutron capture signal is being considered but may effect the long
term stability.

As stated previously, electron antineutrino signal events are isolated using a
coincidence requirement of an outgoing positron followed by a neutron capture.
Background events that mimic these requirements can be divided into two types,
uncorrelated and correlated. The uncorrelated background involves two
independent events that randomly occur in close proximity in time and space.
This type of background can be minimized with low activity passive shielding
and be measured to high precision by swapping the order of the components of
the signal definition. The correlated backgrounds, where both components come
from the same parent event, are more problematic. Examples of this type of
background are two spallation neutrons from the same cosmic ray muon or a
proton recoil produced by a fast neutron. Several methods are available to
mitigate the effects of these correlated backgrounds. Shielding is an
effective method to reduce the cosmic ray rate. For example, the background
rate for a detector at a depth of 300 (600) mwe is 0.2 (0.1) events/ton/day.
One can also create an effectively larger depth by using a high efficiency
veto system to detect and cut out the cosmic-ray muon events that might
initiate these backgrounds. Initial studies indicate that such a system might
reduce the effects of the above background rates by an order of magnitude to a
very low level. The surviving background rate will still need to be measured
but now at only the 25\% level. This can be achieved by using vetoed events to
study distributions and extrapolate into the signal regions.

\section{Examples of possible measurements and comparisons}

From the discussion above, the requirements for a next generation reactor
experiment would include a high power, probably multi-reactor site with the
ability to construct halls and possibly tunnels for the detectors. Hill or
mountains near the site allow more cost effective tunnelling. The ability to
move the far detector to the near site is very desirable and may be crucial to
obtain sensitivities for $\sin^{2}2\theta_{13}$ down to $0.01$.

\begin{figure}
[ptb]
\begin{center}
\includegraphics[
height=2.9196in,
width=3.0242in
]%
{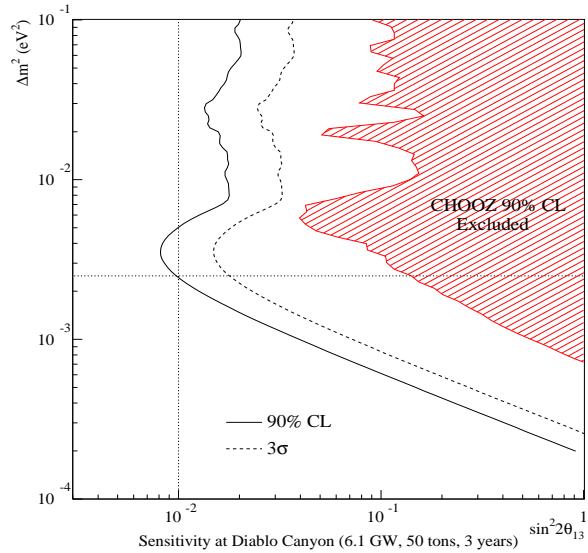}%
\caption{Sensitivity of an oscillation experiment at the Diablo Canyon site
with two 50 ton detectors at 150m and 1200m running for 3 years. The far
detector is at a depth of 600 mwe and the uncertainties in the relative near
to far detector efficiency and far background rate are assumed to be 0.16\%
and 3.5\% respectively.}%
\label{diablo}%
\end{center}
\end{figure}

Many single and two reactor sites exist in the U.S. with average thermal power
in the 3 to 3.5 GW range per reactor. As an example, the Diablo Canyon site
has an average thermal power of 6.1 GW. There are good access roads and nearby
hills that would allow horizontal tunnelling for a far site at 1.2 km with
over 600 mwe of shielding. A three year run of two 50 ton fiducial volume
detectors would provide about 120,000 events in the far detector over a
background of 4900 events and lead to a sensitivity of $\sin^{2}2\theta
_{13}=0.01$ @ 90\% CL for $\Delta m^{2}=2.5\times10^{-3}$ eV$^{2}$ as shown in
Fig. \ref{diablo}.%

%TCIMACRO{\FRAME{ftbpFU}{3.0242in}{2.9196in}{0pt}{\Qcb{Sensitivity of an
%oscillation experiment at the Diablo Canyon site with two 50 ton detectors at
%150m and 1200m running for 3 years. The far detector is at a depth of 600 mwe
%and the uncertainties in the relative near to far detector efficiency and far
%background rate are assumed to be 0.16\% and 3.5\% respectively.}%
%}{\Qlb{diablo}}{diablo_sense.eps}{\special{ language "Scientific Word";
%type "GRAPHIC";  maintain-aspect-ratio TRUE;  display "USEDEF";
%valid_file "F";  width 3.0242in;  height 2.9196in;  depth 0pt;
%original-width 3.6564in;  original-height 3.5284in;  cropleft "0";
%filename 'diablocanyon.eps';file-properties "XNPEU";}}}%
%BeginExpansion
%EndExpansion

Comparisons of various reactor experimental setups are listed in
Table~\ref{tab1}. The examples assume three years of data with two 50 ton
detectors at 150m and 1200m from one or two 3 GW reactors. The cost basis uses
a MiniBooNE cost model for the detector and estimates from a Fermilab NuMI
engineer for the tunnels and halls. The cost inputs were \$5M for the
detector, \$2-3M for a detector hall and \$15-17M for a 1 km tunnel at 300-600
mwe depth. Tunnels were only included for the movable far detector scenarios.

\begin{table}[h]
\tbl{Comparison of various reactor oscillation experiment scenarios.
The $\sin^{2}2\theta_{13}$  column gives the sensitivity at 90\% CL
for $\Delta m^{2}=2.5\times10^{-3}$ eV$^{2}$ The background
contamination is 10,000(5,000) for 300(600) mwe which is assumed to be
measured with an uncertainty of 3.5\% \vspace
*{1pt}}
{\footnotesize
\begin{tabular}
[c]{lllllll}\hline Source & Depth(mwe) & Detector & Events Far &
Rel. Norm Err. & Cost (\$M) & $\sin^{2}2\theta_{13}$ \\\hline One
& \multicolumn{1}{c}{300} & \multicolumn{1}{c}{Fixed} &
\multicolumn{1}{c}{64,000} & \multicolumn{1}{c}{0.008} &
\multicolumn{1}{c}{14} & \multicolumn{1}{c}{0.022}\\
Reactor & \multicolumn{1}{c}{} & \multicolumn{1}{c}{Movable} &
\multicolumn{1}{c}{57,000} & \multicolumn{1}{c}{0.0023} &
\multicolumn{1}{c}{25} & \multicolumn{1}{c}{0.017}\\
& \multicolumn{1}{c}{600} & \multicolumn{1}{c}{Fixed} &
\multicolumn{1}{c}{64,000} & \multicolumn{1}{c}{0.008} &
\multicolumn{1}{c}{16} & \multicolumn{1}{c}{0.020}\\
& \multicolumn{1}{c}{} & \multicolumn{1}{c}{Movable} &
\multicolumn{1}{c}{57,000} & \multicolumn{1}{c}{0.0023} &
\multicolumn{1}{c}{27} & \multicolumn{1}{c}{0.014}\\\hline Two &
\multicolumn{1}{c}{300} & \multicolumn{1}{c}{Fixed} &
\multicolumn{1}{c}{128,000} & \multicolumn{1}{c}{0.008} &
\multicolumn{1}{c}{14} & \multicolumn{1}{c}{0.018}\\
Reactors & \multicolumn{1}{c}{} & \multicolumn{1}{c}{Movable} &
\multicolumn{1}{c}{115,000} & \multicolumn{1}{c}{0.0016} &
\multicolumn{1}{c}{25} & \multicolumn{1}{c}{0.011}\\
& \multicolumn{1}{c}{600} & \multicolumn{1}{c}{Fixed} &
\multicolumn{1}{c}{128,000} & \multicolumn{1}{c}{0.008} &
\multicolumn{1}{c}{16} & \multicolumn{1}{c}{0.017}\\
& \multicolumn{1}{c}{} & \multicolumn{1}{c}{Movable} &
\multicolumn{1}{c}{115,000} & \multicolumn{1}{c}{0.0016} &
\multicolumn{1}{c}{27} & \multicolumn{1}{c}{0.010}\\\hline
\end{tabular}
\label{tab1} } \vspace*{-13pt}\end{table}

\section{Summary}

A next generation reactor experiment could reach a sensitivity to oscillations
with $\sin^{2}2\theta_{13}=0.01$ and $\Delta m^{2}=2.5\times10^{-3}$ eV$^{2}$
at the 90\% CL. The timescales appear reasonable as a complement to the
expected appearance measurements and the cost are not prohibitive. Reactor
experiments can be combined with neutrino only running of off-axis appearance
experiments to isolate CP violation and matter effects. To design a reactor
experiment with 3$\sigma$ sensitivity down to $\sin^{2}2\theta_{13}=0.01$ will
require improvements to the background measurement along with the substantial
betterments of the near to far detector comparison. If a suitable site can be
found and if these improvements can be made, a reactor disappearance
measurement will become a key ingredient to the understanding of neutrino
masses and mixing angles. Several groups around the world are considering this
possibility and expected to submit proposal over the next year.

\end{document}